\documentclass[a4paper,fleqn,usenatbib]{mnras}

\usepackage{newtxtext,newtxmath}

\usepackage[T1]{fontenc}
\usepackage{ae,aecompl}

\usepackage{graphicx}	
\usepackage{amsmath}	
\usepackage{amssymb}	
\usepackage{multirow}

\title[On the Oosterhoff dichotomy in the GB]{On the Oosterhoff dichotomy in the Galactic bulge: II. kinematical distribution}

\author[Prudil et al.]{
Z. Prudil$^{1}$\thanks{E-mail: prudilz@ari.uni-heidelberg.de}, I. D\'ek\'any$^{1}$, E. K. Grebel$^{1}$, M. Catelan$^{2,3}$\thanks{On sabbatical leave at European Southern Observatory, Av. Alonso de C\'ordova 3107, 7630355 Vitacura, Santiago, Chile.}, M. Skarka$^{4,5}$, R. Smolec$^{6}$ \\
$^{1}$ Astronomisches Rechen-Institut, Zentrum f{\"u}r Astronomie der Universit{\"a}t Heidelberg, M{\"o}nchhofstr. 12-14, D-69120 Heidelberg, Germany\\
$^{2}$ Instituto de Astrof{\'i}sica, Pontificia Universidad Cat{\'o}lica de Chile, Av. Vicu{\~n}a Mackenna 4860, 7820436 Macul, Santiago, Chile\\
$^{3}$ Instituto Milenio de Astrof{\'i}sica, Santiago, Chile\\
$^{4}$ Department of Theoretical Physics and Astrophysics, Masaryk University, Kotl\'{a}\v{r}sk\'{a} 2, CZ-311 37, Czech Republic\\
$^{5}$ Astronomical Institute, Czech Academy of Sciences, Fri\v{c}ova 298, CZ-251 65, Ond\v{r}ejov, Czech Republic\\
$^{6}$ Nicolaus Copernicus Astronomical Center, Polish Academy of Sciences, ul. Bartycka 18, 00-716 Warszawa, Poland\\
\\
}

\date{Accepted XXX. Received YYY; in original form ZZZ}

\pubyear{2017}

\begin{document}

\label{firstpage}
\pagerange{\pageref{firstpage}--\pageref{lastpage}}
\maketitle

\begin{abstract}
We present a kinematical study of RR~Lyrae stars associated with two Oosterhoff groups in the Galactic bulge. We used data published in the first paper of the series, plus proper motions from the {\it Gaia} Data Release 2, and radial velocities from the literature. A 6D kinematical and spatial solution was obtained for 429 RR~Lyrae stars. We use a model of the Galactic gravitational potential to infer stellar orbits. We did not find a difference between the Oosterhoff groups in the individual components of the space velocity. We report that \textit{foreground} and \textit{background} stars with respect to the Galactic bulge stand out in the mean $V$ velocity component, which we interpret as a sign of the Galactic rotation. The movement of the studied stars in the central region of the Galactic bulge is consistent with random motions expected for a classical bulge component. From the orbital integration, we estimate that 8\,\% of the RR~Lyrae stars are halo interlopers currently located in the Galactic bulge. The majority of the stars' orbits are within a 3\,kpc radius from the Galactic bulge. The fraction of Oosterhoff\,II stars increases with increasing Galactic latitude, as well as towards longer orbital periods. We found several RR~Lyrae stars with high space velocities, one of which has an extremely long orbital period of $\sim$1\,Gyr. We conclude that based on their kinematics, the vast majority of the stars in our sample do not seem to contribute to the boxy/peanut component of the Galactic bulge.
\end{abstract}




\section{Introduction}
According to the current paradigm, galactic bulges form in two main ways: the bar-like pseudobulges through instabilities of the disk, and the spheroidal (classical) bulges through early mergers. The pseudobulges are characterized by cylindrical rotation while random motion dominates the classical bulges \citep[e.g.,][]{Wyse1997,Kormendy2004,Barbuy2018}. 

The Milky Way (MW) bulge has a pronounced boxy/peanut bulge component recognized already in images from the {\it COBE} satellite \citep{Weiland1994,Dwek1995} and later spatially studied using red clump giants by, e.g., \citet{McWilliam2010} and \citet{Wegg2013}, among many others. Recent studies suggest that a pseudo- and a classical bulge component (one from secular formation, the other via mergers) can co-exist \citep{Obreja2013}, but whether our own Galaxy's bulge contains both components is still a matter of a debate \citep[e.g.,][]{Zoccali2008}, for instance, due to the possible vertical metallicity gradient \citep[e.g.,][]{Zoccali2008,Ness2013b}. Kinematically, the Galactic bulge has been studied mainly using red clump stars \citep[The Abundances and Radial velocity Galactic Origins Survey - ARGOS,][]{Freeman2013,Ness2013a} and using M~giants \citep[Bulge Radial Velocity Assay - BRAVA,][]{Kunder2012}. Both surveys targeted mainly the intermediate-age population of the Galactic bulge, and supported the pseudo-bulge morphology.

In this work, we will focus on tracers of the oldest bulge population, as represented by the RR~Lyrae variables. They are short-period pulsating stars burning helium in their cores and are associated with Population II \citep{Catelan2015Book}. They are ideal probes of the old stellar systems in the Local Group, for instance through their period-luminosity dependence \citep{Alcock1997,Sesar2013,Dekany2013}. They have been proven a useful tool in many branches of astrophysics, ranging from in-depth studies of individual stars \citep[e.g.,][]{Guggenberger2012} to studies of Galactic structure and evolution \citep{Dekany2013,Dekany2018,Pietrukowicz2015}. 

Despite their abundance in the MW and in the Magellanic system, there are still some unanswered questions related to RR~Lyrae stars, e.g. the origin of the Oosterhoff dichotomy \citep{Oosterhoff1939}. The Oosterhoff dichotomy divides MW globular clusters containing RR~Lyrae stars into two main groups, traditionally labeled Oosterhoff I (Oo\,I) and Oosterhoff II (Oo\,II), based on the pulsation properties of the associated RR~Lyrae variables. The Oosterhoff type I globular clusters contain RR~Lyrae stars with short pulsation periods and are more metal-rich in comparison with globular clusters of Oosterhoff type II, which are metal-poor and contain RR~Lyrae variables with longer pulsation periods. At the same time, GCs in the Magellanic Clouds and nearby dwarf galaxies do not fit into either of the two Oo groups in general, which connects the Oo dichotomy to the MW itself. In addition, two globular clusters located in the Galactic bulge, NGC 6441 and NGC 6388, constitute a third Oo group of their own \citep[Oo\,III]{Pritzl2000}. We point the interested reader to \citet{Catelan2009} and \citet{Catelan2015Book} for comprehensive reviews of the Oosterhoff dichotomy.

The RR~Lyrae stars have been used for kinematical studies in the past, e.g. \citet{Layden1996} showed that the metal-rich ones ([Fe/H]$ > -1.0$\,dex) in the nearby Galactic field belong to the thick disk population while the rest probably originate from the halo. In addition, recent kinematic studies targeted RR~Lyrae stars in the Galactic bulge \citep{Kunder2015,Kunder2016,Kunder2018} concluding that they are part of a classical bulge or metal-rich inner halo/bulge component with a random orbital distribution. 

In the first paper of this series \citep{Prudil2019}, we studied the spatial distribution of RR~Lyrae stars in the Galactic bulge in the context of the Oosterhoff dichotomy. We found that $\approx$~25\,\% of the RR~Lyrae variables in the Galactic bulge are associated with the Oosterhoff II group. Pulsators of the Oosterhoff II population are on average more massive, brighter, and cooler than their Oosterhoff I counterparts. We did not find any link between the individual Oosterhoff populations and the Galactic bar. 

In this paper, we present a kinematical study of the Oosterhoff groups in the Galactic bulge based on several hundred RR Lyrae variables used in the previous work. In Sec.~\ref{sec:dataSample}, we describe the selection criteria and basic statistics of our sample. In Sec.~\ref{sec:Kinematika_oosy} we outline the process of calculating kinematical properties and discuss sources of possible contamination in our sample. Section \ref{sec:Orbits} discusses the orbital integration of our sample using a gravitational potential model of the MW and their orbital parameters. A few particularly interesting objects are addressed in greater detail. In Sec.~\ref{sec:Conclus} we summarize our findings.

\section{Data sample}  \label{sec:dataSample}

We used the fundamental-mode RR~Lyrae stars located toward the Galactic bulge with estimated metallicities, distances, and Oosterhoff status from \citet{Prudil2019}. The distances and metallicities were obtained using photometric data from the fourth phase of the Optical Gravitational Lensing Experiment \citep[hereafter OGLE-IV,][]{Udalski2015,Soszynski2017} and the VISTA Variables in the V\'ia L\'actea survey \citep[VVV,][]{Minniti2010}. 

In this study, we will use data from the second {\it Gaia} data release, the {\it Gaia} DR2 \citep{Gaia2016,Gaia2018}. It contains multiband photometry for 1.7 billion sources together with parallaxes and proper motions for 1.3 billion sources. From our sample, we found a match for 7\,785 objects with proper motions and parallax measurements, thus nearly for all selected stars.

Some of the cross-matched variables have unreliable proper motions (with an error in proper motion higher than the proper motion itself). Thus, we removed stars in which the error in proper motion exceeds 50\,\% of its proper motion value. The majority of the removed pulsators have negligible proper motion either in R.A. or Dec. In the end, 6\,542 variables were considered for further analysis. We investigated whether stars with different Oosterhoff classifications have different distributions of proper motion in either direction using the two-sample Kolmogorov-Smirnov test. This statistical test returns the probability $p_{\rm KS}$ of obtaining our sample under the null hypothesis that they have identical distributions. The null hypothesis is usually rejected if $p_{\rm KS}<0.05$. Our test resulted in $p_{\rm KS}=0.33$, thus we conclude that there is no significant difference between the proper motion distributions of the tho Oosterhoff groups.

If we assume that RR~Lyrae stars trace the spheroidal component of the Galactic bulge they should belong to its kinematically hot component with a minor rotation component and large velocity dispersion. The tangential velocities $v^{t}$ can be used to test such an assumption through a simple equation, $v^{t} = 4.74 \cdot \mu \cdot d$, with the proper motion $\mu$ in mas\,yr$^{-1}$ and distance $d$ in kpc. For individual proper motion components we a get dispersion $\sigma v^{t}_{\alpha\ast} = 125$\,km\,s$^{-1}$ and $\sigma v^{t}_{\delta} = 123$\,km\,s$^{-1}$. Therefore, the tangential velocity ellipsoids are fairly symmetric and support the assumption about the RR~Lyrae stars being part of the kinematically hot component of the Galactic bulge. This conclusion is in agreement with a study using the Galactic bulge Type II Cepheids \citep{Braga2018}.

\begin{figure}
\includegraphics[width=\columnwidth]{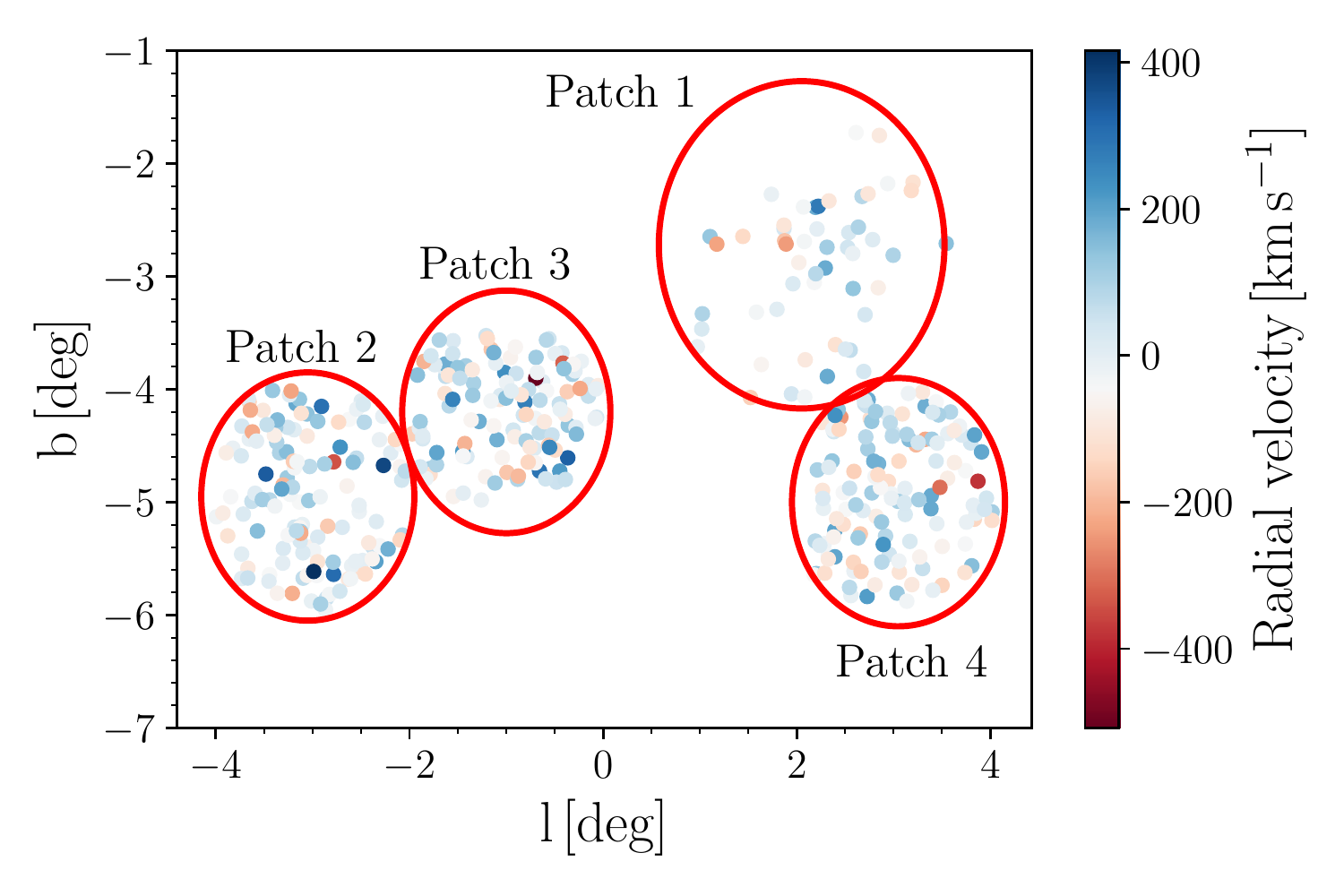}
\caption{The spatial distribution of our sample variables in Galactic coordinates with color-coding representing the measured radial velocity.}
\label{fig:radialky}
\end{figure}

The full information of space velocities can hint at possible differences between stars if they differ in origin. Therefore, for the purpose of this paper, we cross-matched stars with distances and proper motions with the published catalog of \citet{Kunder2016}, consisting of radial velocities for 947 RR~Lyrae pulsators in the Galactic bulge. The cross-matched sample consists of 429 variables (313 Oo\,I and 116 Oo\,II type). Figure \ref{fig:radialky} shows the distribution of radial velocities for our sample of variables in Galactic coordinates in four observed patches. The radial velocities for the Oo\,I variables vary from $-381$\,km\,s$^{-1}$ to $337$\,km\,s$^{-1}$ with a median of $8$\,km\,s$^{-1}$. On the other hand, Oo\,II stars have a broader range of values (from $-508$\,km\,s$^{-1}$ to $416$\,km\,s$^{-1}$) with a median of $12.5$\,km\,s$^{-1}$. The K-S test of the radial velocity distribution of both Oosterhoff populations yields the following values: $\text{D}=0.100$ and $p_{\rm KS}=0.344$. Therefore, the radial velocities seem to be drawn from the same distribution, and there is no difference between both groups.  

We note that the selected sample of radial velocities does not contain individual errors. Therefore, we decided to assume a constant error value of 10\,km\,s$^{-1}$ for each star. Our assumption is based on errors in the radial velocities of variables in NGC 6441 \citep{Kunder2018}, where 40 RR~Lyrae stars have been observed with the same instrument, similar exposure time (from 0.5 to 2 hours), and the same resolution as for the pulsators in \cite{Kunder2016}. Their errors are on average $\pm$7.5\,km\,s$^{-1}$ and only 3 stars have higher errors than 10\,km\,s$^{-1}$. Thus, on average, we slightly overestimate the velocity errors.

\section{Kinematics of the Oosterhoff populations in the Galactic bulge} \label{sec:Kinematika_oosy}

In this section, we study the kinematics of the different Oosterhoff groups in the Galactic bulge. 

The components of the space velocity were obtained following \citet{Johnson1987} in the $U$, $V$, and $W$ Galactic Cartesian system. We assumed a left-handed system with positive $U$ towards the Galactic anticenter, $V$ in the direction of Galactic rotation, and positive $W$ in the direction to the Galactic north pole. We adopted the coordinates of the Galactic north celestial pole based on the coordinates provided by the Hipparcos satellite \citep{ESA1997}:
\begin{gather}
\alpha_{GP} = 192.85948^{\circ}, \hspace{1cm}
\delta_{GP} = 27.12825^{\circ},\\ 
l_{GP} = 122.93192^{\circ},
\end{gather}  
where $\alpha_{GP}$ and $\delta_{GP}$ represent the coordinates of the Galactic north celestial pole and $l_{GP}$ stands for the position angle between the Galactic north celestial pole and zero Galactic longitude. For the peculiar solar motion we assumed $\left(U_{\odot},V_{\odot},W_{\odot} \right) = \left(-14.1\pm0.57, 12.24\pm0.47, 7.25\pm0.37\right)$\,km\,s$^{-1}$ \citep{SCHONRICH2010,SCHONRICH2012} and a local standard of rest velocity $v_{LSR}=220$\,km\,s$^{-1}$ \citep{Kerr1986,Bovy2012}. 

The true space velocity $sv$ can be obtained through the simple equation:
\begin{equation}
sv = \sqrt{U^{2} + V^{2} + W^{2}}.
\end{equation}
The median values and dispersions of the individual $sv$ components are listed in Tab.~\ref{tab:statisOO}. In Fig.~\ref{fig:Elipsoids-cont} (top and side insets) we display the distributions of the space velocity and its Cartesian components. The errors for the individual velocity components were obtained through 10000 iterations of the Monte Carlo error sampling where we added Gaussian random noise to the solar motion, distances, radial velocities, and proper motions with $\sigma$ made equal to their uncertainties. From these four panels, we see that both Oosterhoff groups are centered at 0\,km\,s$^{-1}$ and that the distributions are fairly similar for the velocities in the Galactic plane $U$ and $V$. A small systematic difference between the two groups can be seen in the $W$ component, where on the positive side we see two peaks (roughly at 58\,km\,s$^{-1}$ and 187\,km\,s$^{-1}$, marked with black arrows in Fig.~\ref{fig:Elipsoids-cont}). Closer examination of the spatial and kinematical properties of the stars responsible for this deviation did not show any sign of clustering or any preferential direction of velocities. The distribution of space velocities is depicted in the bottom-right panel of Fig.~\ref{fig:Elipsoids-cont}. As in the case of individual components, we do not see any major differences between the two Oosterhoff groups. The green dashed line represents the true space velocity found by \citet{Kunder2015} for one of the high-velocity RR~Lyrae stars in the Galactic bulge. From our sample, six stars have higher velocities than the variable found by \citet{Kunder2015}; two of these stars are discussed in detail in Sec.~\ref{subsec:Pecul}.

In order to test whether the distribution of space velocity components differs between Oosterhoff groups, we used the crossmatch test (CM test) for multivariate distributions \citep{Rosenbaum2005} using the package implementation \texttt{crossmatch} in \texttt{R}. The CM test uses the distances between observations for comparison. In our case, we calculated Mahalanobis distances (distance between a point and a distribution) instead of the commonly used Euclidian distances since our data are highly correlated. The CM test works only with distributions having an even number of values, therefore for our sample, we calculated the $p$-value for the entire distribution while always leaving one star out. In the end, we ran this calculation 429 times and calculated the median of the resulting $p$-values. The comparison for the velocity components $U$, $V$, and $W$ in both Oosterhoff groups yields $p_{\rm CM}=0.32$ following the same considerations as in Sec.~\ref{sec:dataSample}, therefore the distribution of the velocity components for the two Oosterhoff groups in the Galactic bulge is fairly similar. In addition, we calculated the same test for the distribution of the Galactocentric Cartesian coordinates ($x$, $y$, and $z$) for our sample, and the resulting $p$-value, $p_{\rm CM}=0.71$, suggests that the distribution of both Oosterhoff groups in the Galactic bulge is very similar.

\begin{table*}
\caption{Statistical properties of the velocity distributions displayed in Fig.~\ref{fig:Elipsoids-cont}. The first column lists both Oosterhoff groups, while individual velocities are listed in columns 2, 4, 6, and 8 with respect to the two Oosterhoff populations. Columns 3, 5, 7, and 9 list the dispersion in the components of $sv$ and in $sv$ proper, respectively.}
\label{tab:statisOO}
\begin{tabular}{lcccccccc}
\hline  
                      & $U$\,[km\,s$^{-1}$]    & $\sigma^{U}$\,[km\,s$^{-1}$]  & $V$\,[km\,s$^{-1}$] & $\sigma^{V}$\,[km\,s$^{-1}$]  & $W$\,[km\,s$^{-1}$]   & $\sigma^{W}$\,[km\,s$^{-1}$] & $sv$\,[km\,s$^{-1}$] & $\sigma^{sv}$\,[km\,s$^{-1}$] \\ \hline 
Oo\,I  & -20 & 118 & -22 & 123 & 10 & 121 & 192 & 89 \\
Oo\,II & -35 & 126 & -25 & 128 & 15 & 129 & 203 & 99 \\     
\hline
\end{tabular}
\end{table*}

\subsection{Possible contamination}

\begin{figure*}
\includegraphics[width=\columnwidth]{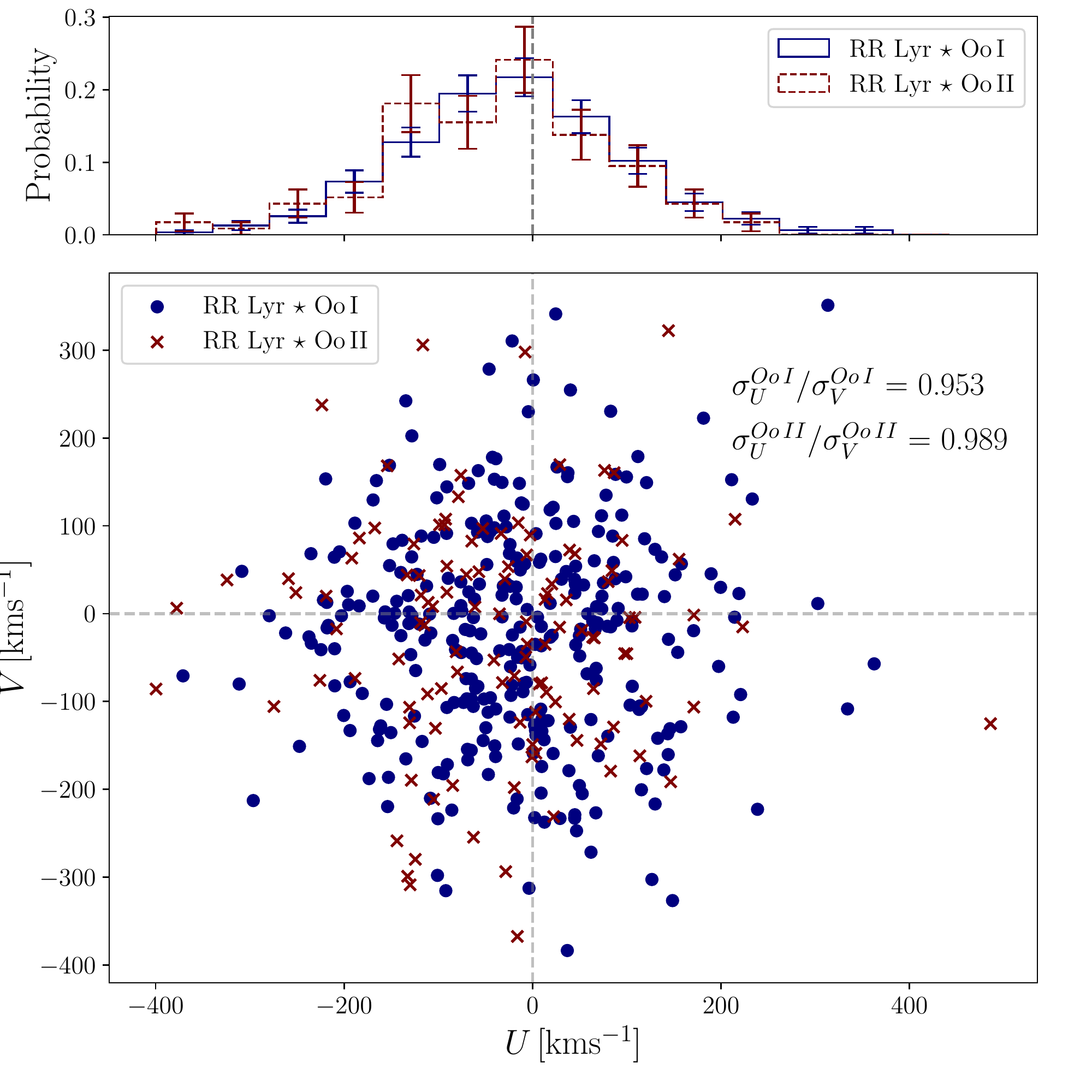}
\includegraphics[width=\columnwidth]{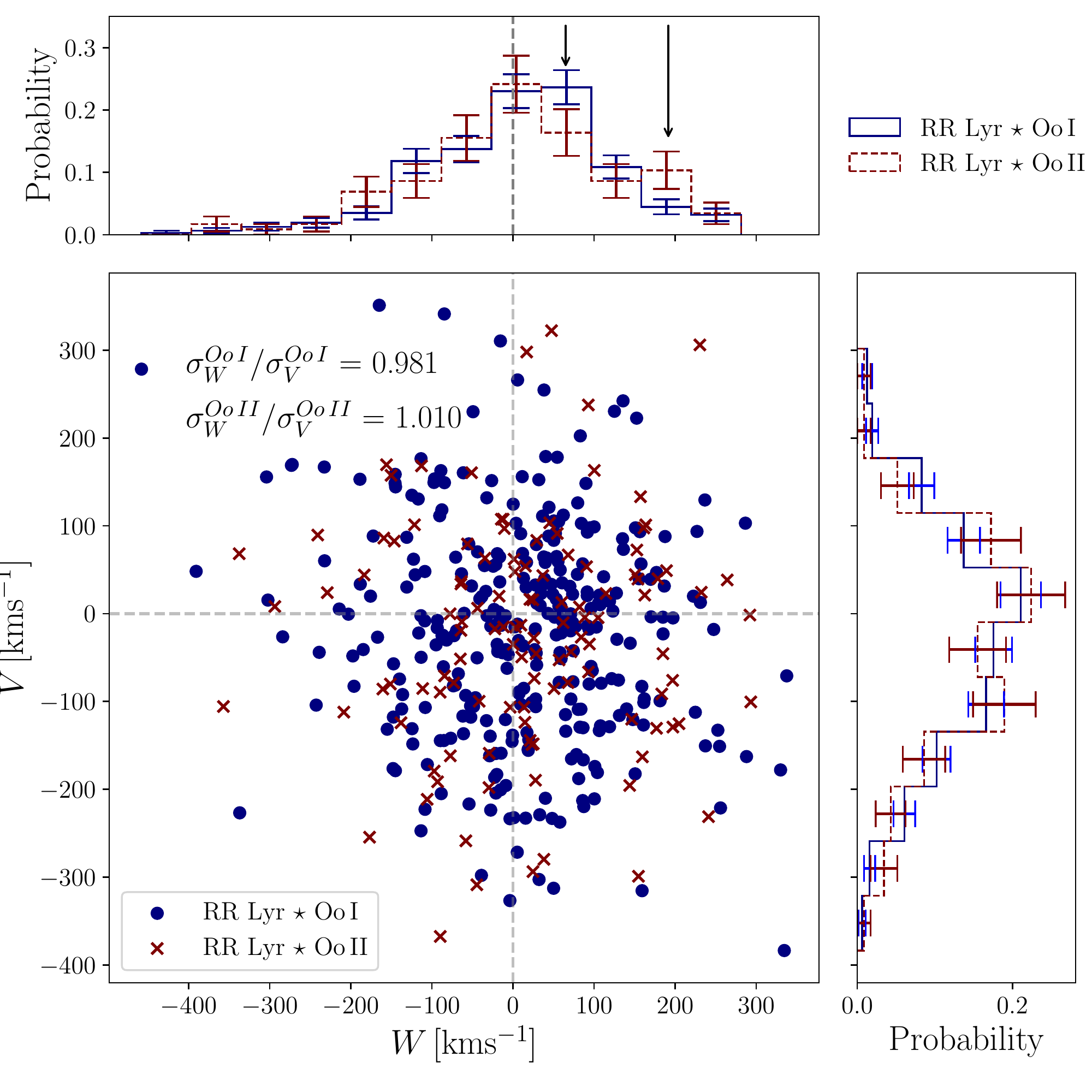} \\ \vspace{-0.25cm}
\includegraphics[width=\columnwidth]{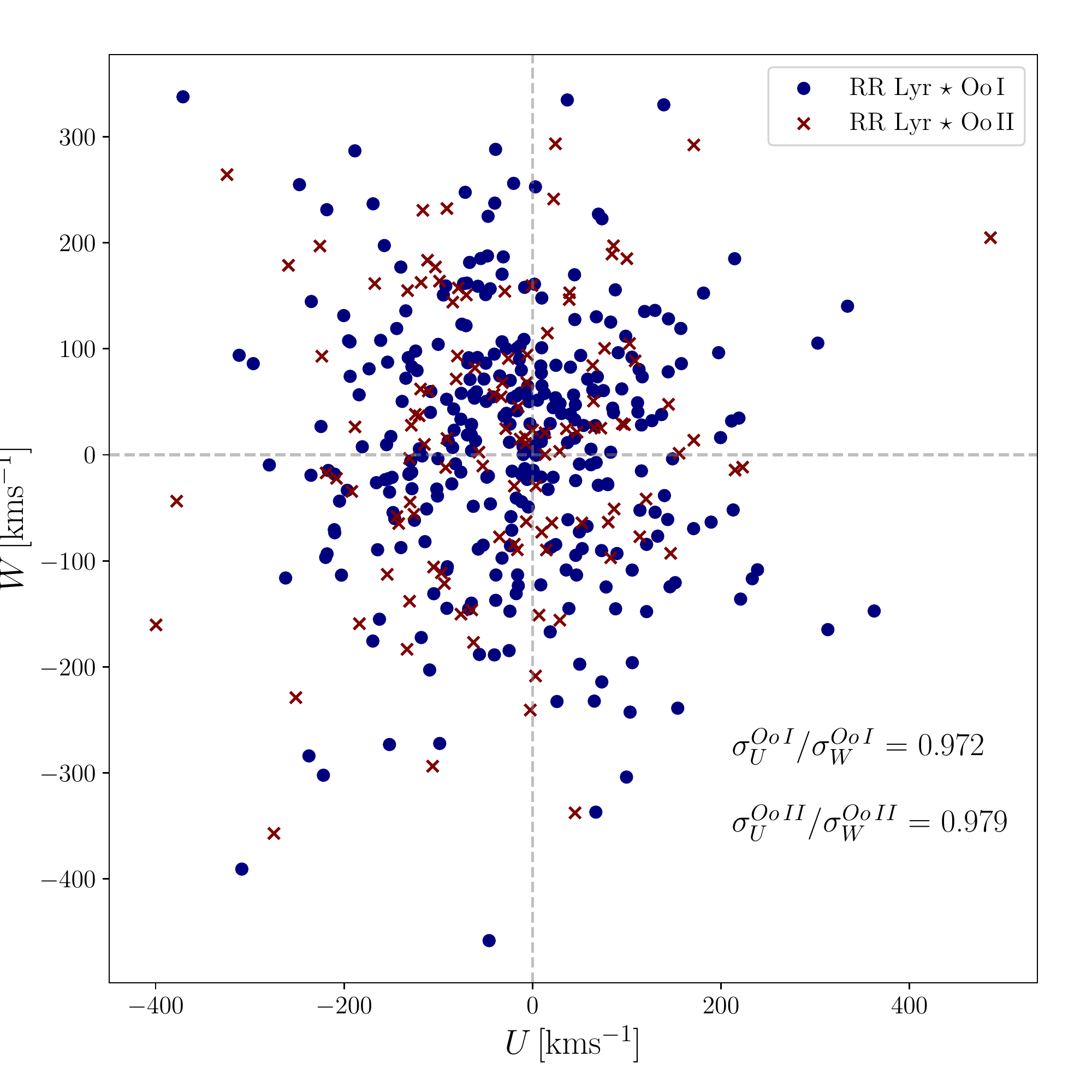}
\includegraphics[width=\columnwidth]{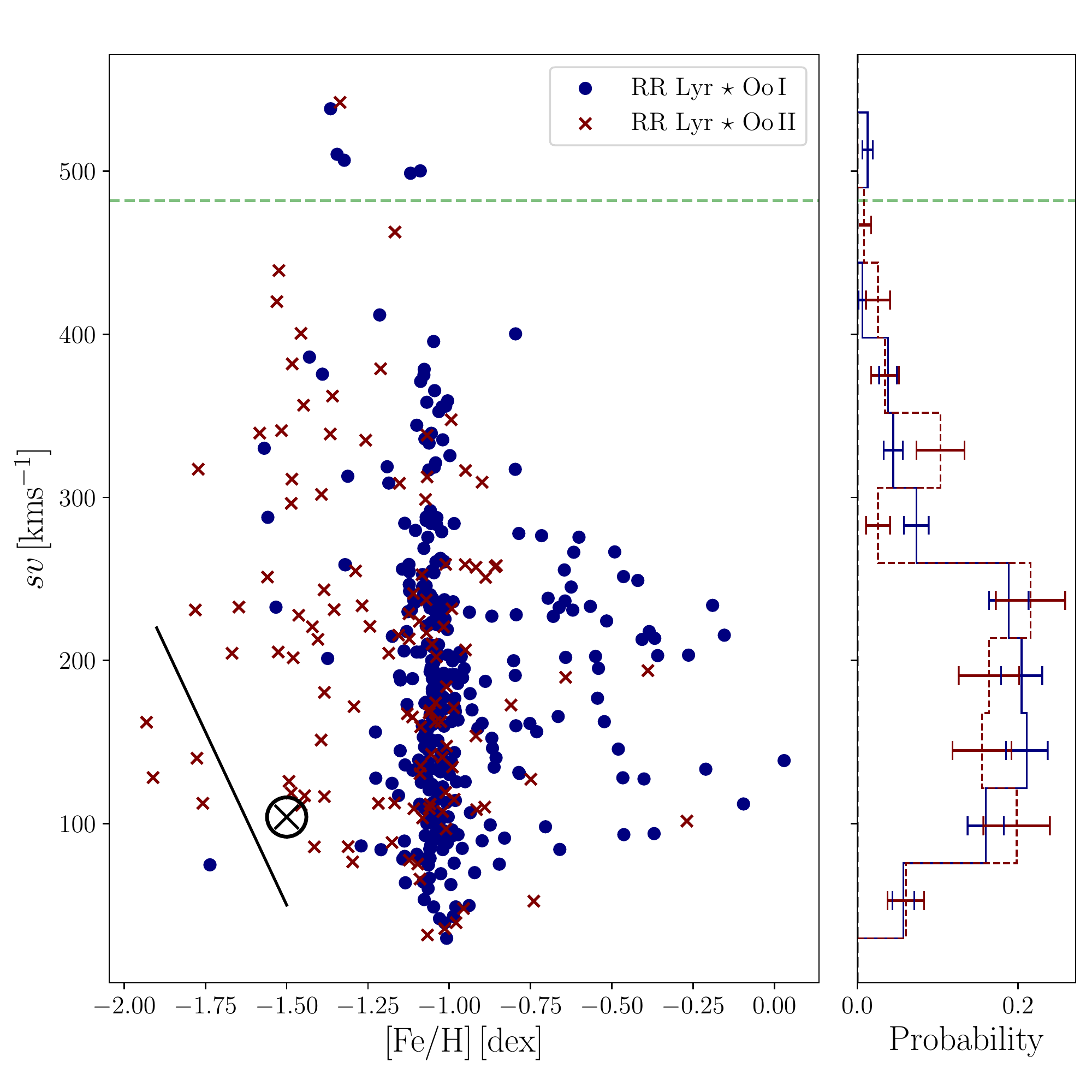}
\caption{The dependences of individual components of the $sv$ with respect to each other (top left and right and bottom left panels) and $sv$ vs. [Fe/H] dependence (bottom right panel). The blue points and red crosses represent the Oo\,I and Oo\,II population, respectively. The dashed green line in the bottom right panel stands for the high-velocity bulge RR~Lyrae stars ($sv=482\,{\rm km\,s}^{-1}$) found by \citet{Kunder2015}. The black line in the same panel separates 5 variables with very low metallicity and $sv$ values. In addition, the black crossed circle represents the average position of RR~Lyrae stars in NGC 6441 with respect to their $sv$ and photometrically determined [Fe/H].\protect\footnotemark}
\label{fig:Elipsoids-cont}
\end{figure*}

Our sample might contain some stars from the Galactic disk and/or halo that would cause contamination. In Fig.~\ref{fig:Elipsoids-cont} we plotted dependences of individual velocities, calculated their velocity ellipsoids, and displayed the distribution of the $sv$ with respect to the photometric metallicity. The velocity ellipsoids are quite \textit{round}, with ratios of dispersion varying around 0.9 $\sim$ 1.0 (see the annotations in the individual panels). 

In the $sv$ vs. metallicity plane, we see several interesting features. The upper envelope of the space velocities of RR~Lyrae stars with [Fe/H] $\sim$-1.0\,dex seems to monotonically approach 400\,km\,s$^{-1}$ as we move towards more metal-poor variables. The bulk of the studied variables have velocities below 400\,km\,s$^{-1}$ with only 11 stars exceeding this value. The velocities of these 11 RR~Lyrae increase with decreasing metallicity, which suggests an origin outside the Galactic bulge (i.e., in the metal-poor halo).  

In addition, five stars in the bottom-left corner of this panel exhibit very low metallicity and $sv$ values. This behaviour is in contrast with the general trend in our sample, which shows higher velocities for lower metallicities. These five stars have rather low radial velocities (average 19\,km\,s$^{-1}$), and preferentially longer pulsation periods (average 0.75\,days). Two out of these five stars fall into the region of the Oo\,III group for the globular cluster NGC 6441 in the period-amplitude diagram \citep[see fig.~5 in][]{Prudil2019}. If they belonged to that globular cluster, their metallicity from the photometric data would not agree with the metallicity determined from spectroscopic measurements and thus their distance, determined on basis of photometric metallicity, would be incorrect. 

Some of the globular clusters in the Galactic bulge seem to be dissolving \citep[e.g., NGC 6441,][]{Kunder2018}. Therefore, some of the stars in our sample might have been stripped from NGC 6441 in the past and have mixed with the bulge field population of RR~Lyrae stars. In addition, one of these stars is a candidate for the double-mode variables found by \citet{Smolec2016}, which, based on stellar pulsation models, have a high metallicity despite the low metallicity determined using photometric data. After a careful analysis of the Fourier spectra of the remaining four stars, none of them shows any sign of a double-mode behavior. 

In addition, the calculated orbits of these 5 stars show that 4 of them stay within the $\approx$ 3.5\,kpc radius of the Galactic bulge and do not deviate more than 1\,kpc from the Galactic plane during their orbits (for details on orbital integration see Sec.~\ref{sec:Orbits}). Thus, they may be interesting candidates for spectroscopic follow-up studies, since they stand out in comparison with other bulge RR~Lyrae stars due to their uncertain origin.

Overall, we do not see a major disk/halo contribution to our sample. Further examination with orbital solutions of the studied variables can be found in Sec.~\ref{sec:Orbits}.

\footnotetext{We note that the calculated photometric metallicity does not agree with the metallicity determined through spectroscopy \citep{Clementini2005}. We use this metallicity to emphasize our point that slowly-moving RR~Lyrae stars with large photometric metallicity may actually belong to the Oo\,III group \citep{Pritzl2000}.}

\subsection{Distribution of velocities in space}

\begin{figure*}
\includegraphics[width=524pt]{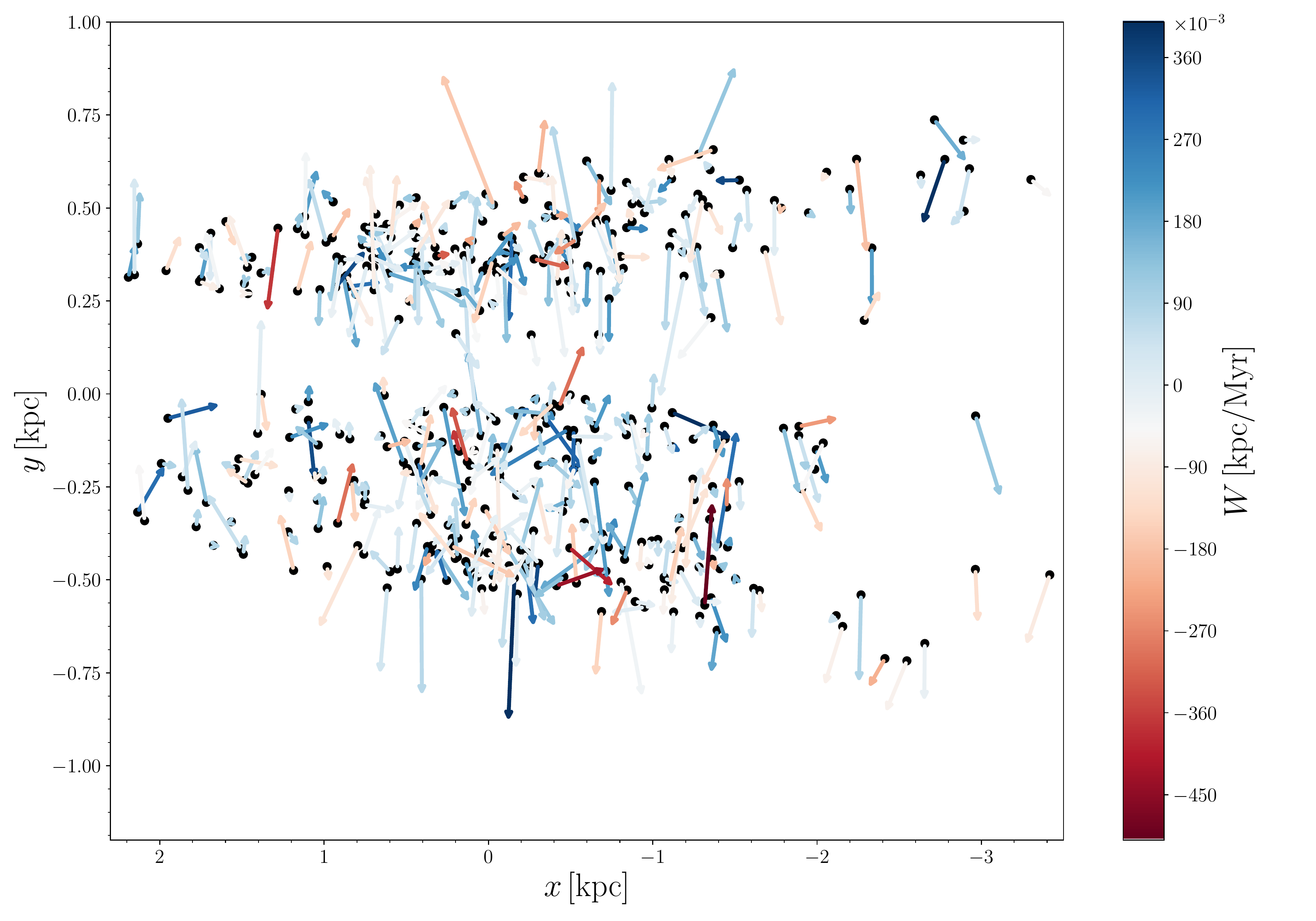}
\caption{The spatial distribution of variables with calculated space velocities (black dots). The arrows represent the directions and values of the velocities $U$ ($\dot{x}$\,[kpc/Myr]) and $V$ ($\dot{y}$\,[kpc/Myr]) with the third component $W$ ($\dot{z}$\,[kpc/Myr]) depicted using the color-code indicated on the right.}
\label{fig:SpatUVW}
\end{figure*}

Fig.~\ref{fig:SpatUVW} shows the spatial distribution of our stars with components of the space velocity. The space velocity components for stars in the central region of the Galactic bulge do not show any preferential direction and resemble random motions. We do not find any clumps nor groups of stars with similar location and following a preferential direction. We searched for stars with similar proper motions and radial velocities (difference in proper motion $<$0.1\,mas\,yr$^{-1}$ and radial velocity $<$20\,km\,s$^{-1}$). For possible candidates, we calculated their orbits (see Sec.~\ref{sec:Orbits}) to further investigate their possible link. In the end, we did not find any connection between the stars in our sample.  

In addition, we used a catalog for the Milky Way globular clusters, containing their distance, average proper motions, and radial velocities \citep{Vasiliev2019}. We inspected stars whose kinematic elements (proper motion and radial velocity) match the kinematic features of globular clusters, using the same criteria as in the search for groups of stars with similar kinematics. Using orbital integration (see Sec.~\ref{sec:Orbits}) we then integrated the orbit of selected candidates (the variables and globular clusters) forward and back in time. We did not find any sign of an association between the sample stars and retrieved globular clusters.

\begin{figure}
\includegraphics[width=\columnwidth]{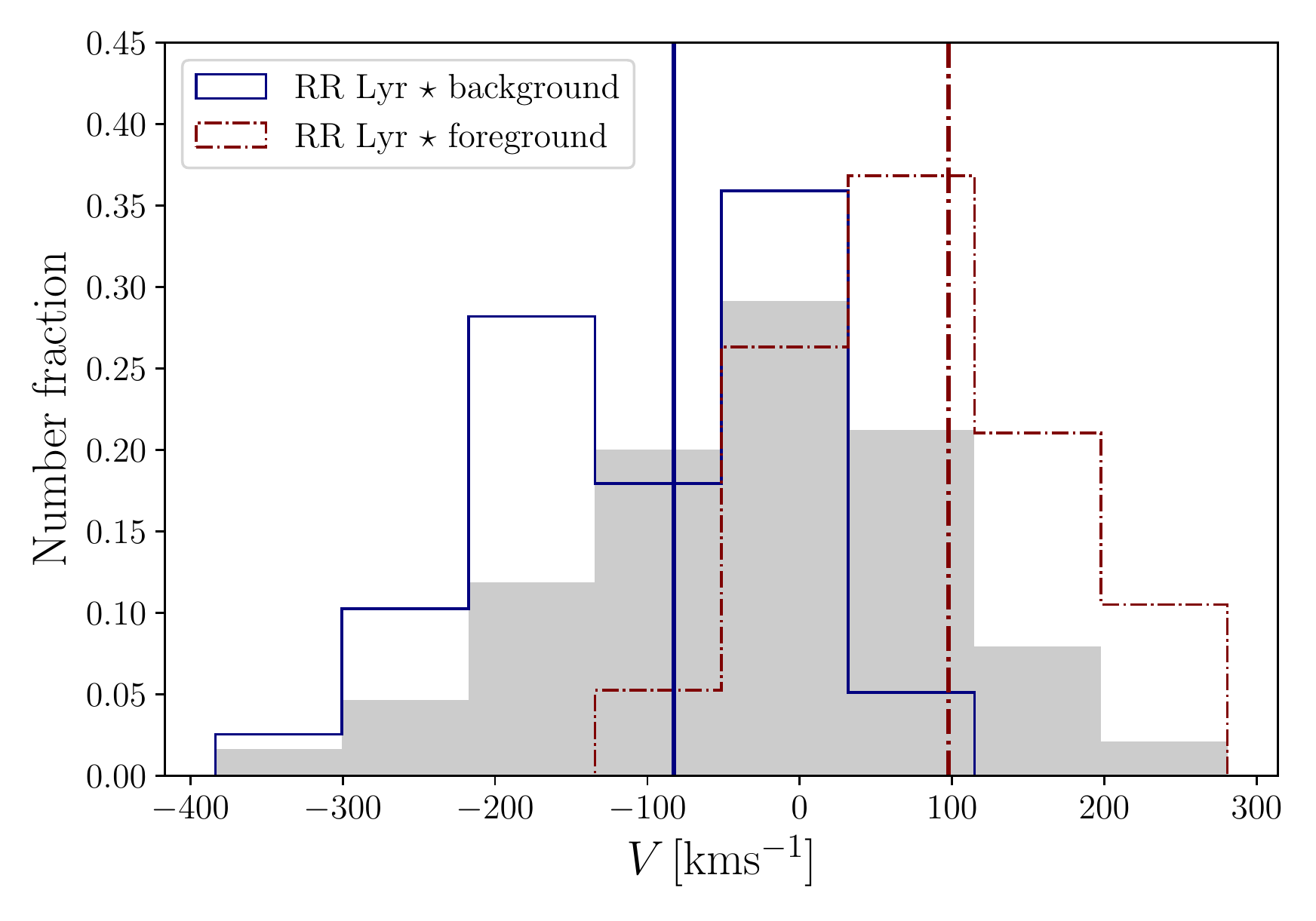}
\caption{The distribution of the $V$ velocity components for the studied stars (grey histogram). The red and blue histograms represent the distribution of RR~Lyrae stars in the foreground and background of the Galactic bulge, respectively, with the same color-coding for vertical lines denoting the median velocities for a given region.}
\label{fig:SpatRotation}
\end{figure}

We do observe an opposite general direction of the space velocity component $V$ for the foreground ($x > 1.75$\,kpc) and background ($x < -1.75$\,kpc) of the Galactic bulge (see Fig.~\ref{fig:SpatRotation}). Regardless of the Oosterhoff group in the \textit{foreground}, the majority of variables move in the direction of the Galactic rotation, with a median velocity $V=97$\,km\,s$^{-1}$. In the \textit{background} of the Galactic bulge, irrespective of the Oosterhoff group, the majority of pulsators move in the opposite direction, with a median velocity of $V=-83$\,km\,s$^{-1}$. This is a clear sign of the Galactic rotation in the front of and behind the Galactic bulge. Neither of the Oosterhoff groups seems to be dominant in this motion, they both contribute to this pattern equally. We note that, for RR~Lyrae stars in the background of the Galactic bulge, we see in Fig.~\ref{fig:SpatRotation} two peaks and a small dip between them. Using the algorithm \texttt{Skinny-dip}\footnote{\url{http://www.kdd.org/kdd2016/subtopic/view/skinny-dip-clustering-in-a-sea-of-noise}}, which is based on Hartigan's dip test of unimodality \citep{Hartigan1985}, we got a $p$-value$=0.48$ following the same considerations as in Sec.~\ref{sec:dataSample}, which supports a unimodal distribution of $V$ velocities behind the Galactic bulge. For these stars, assuming they rotate with the bar/disk, their Galactocentric velocity ($V_{GC}$) should increase with increasing Galactic longitude -- $l$ \citep[see fig.~3 in][]{Ness2013a}. For selected variables associated with the \textit{foreground} and \textit{background} we do not observe such an effect, most likely due to such factors as, e.g., low width in $l$ (for stars around 0\,deg in $l$ the $V_{GC}$ varies around 0\,km\,s$^{-1}$) and the fact that our sample contains only a few dozen objects.

\section{Orbits of the Oosterhoff populations} \label{sec:Orbits}

To compare the orbits and orbital parameters of both Oosterhoff groups we used the \texttt{galpy}\footnote{\url{http://github.com/jobovy/galpy}} library \citep{galpy2015} to integrate orbits of individual stars. We utilized the axisymmetric model for the MW Galactic potential in \texttt{galpy}, \texttt{MWPotential2014}, which is composed of a power-law density profile for spherical bulges with exponential cut-off, Miyamoto-Nagai disk \citep{MiyamotoNagai1975}, and Navarro-Frenk-White halo potential \citep{Navarro1997}. We also added the \texttt{KeplerPotential} for the supermassive black hole in the MW center with a mass $4 \cdot 10^{6}\,M_{\odot}$ \citep{Gillessen2009} to account for close encounters, and the \texttt{DehnenBarPotential} for the Galactic bar \citep{Dehnen2000}. For all of the following calculations, we assumed a rotation velocity of 220\,km\,s$^{-1}$ \citep{Kerr1986,Bovy2012}, a distance to the Galactic center of 8.2\,kpc \citep{BHG2016}, and the same peculiar solar motion as in Sec.~\ref{sec:Kinematika_oosy}, and we integrated forward in time up to 1\,Gyr.

The \texttt{galpy} software allows one to calculate parameters of individual orbits, including e.g. eccentricity $e$, maximum height above or below the Galactic plane $z_{\text{max}}$, and peri- and apocenter of the orbit, $r_{\text{peri}}$ and $r_{\text{apo}}$, respectively. For each star, we ran a Monte Carlo simulation. We calculated 200 orbital solutions in which we varied the initial conditions (proper motions, radial velocities, and distances) within their uncertainties assuming that their distribution is Gaussian. From the resulting distribution, we calculated the median orbital parameters for both Oosterhoff groups. The results are listed in Table~\ref{tab:ComponentsGALPY}. In order to test the differences between the two Oosterhoff groups we used the CM test (see Section\,\ref{sec:Kinematika_oosy}). For the entire distribution of orbital parameters $z_{\text{max}}$, $e$, $r_{\text{peri}}$, and $r_{\text{apo}}$, the CM test yields $p_{\rm CM}=0.68$, there is no statistical evidence that the two distributions are not identical.

\begin{table}
\centering
\caption{Median values of calculated parameter orbits for the individual Oosterhoff groups. Column 1 lists the studied components, colums 2 and 3 list values for each Oosterhoff group.}
\label{tab:ComponentsGALPY}
\begin{tabular}{lcccc}
\\ \hline
 & Oo\,I & Oo\,II \\ \hline
z$_{\text{max}}$\,[kpc] & 1.136 $\pm$ 0.295 & 1.280 $\pm$ 0.363 \\
$e$ & 0.855 $\pm$ 0.052 & 0.817 $\pm$ 0.078 \\
$r_{\text{apo}}$\,[kpc] & 2.212 $\pm$ 0.641 & 2.416 $\pm$ 0.886 \\
$r_{\text{peri}}$\,[kpc] & 0.211 $\pm$ 0.122 & 0.304 $\pm$ 0.176 \\ \hline
\end{tabular}
\end{table}

\begin{figure}
\includegraphics[width=\columnwidth]{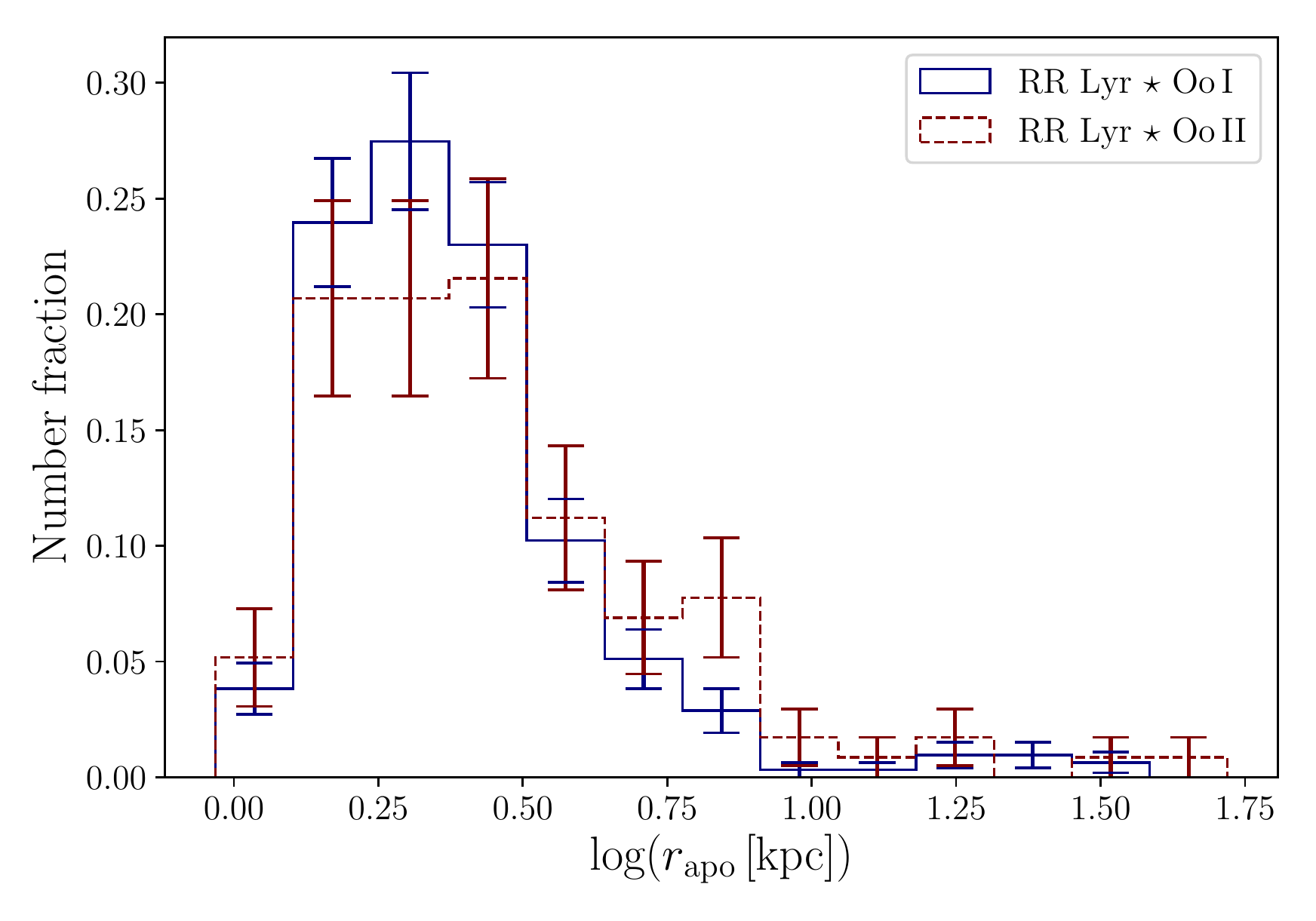} \\
\caption{The distribution of apocentric distances for the studied variables. The blue and red histograms represent the Oo\,I and Oo\,II groups, respectively.} 
\label{fig:ApocHist}
\end{figure}

Figure~\ref{fig:ApocHist} depicts the distribution of apocentric distances for the studied RR~Lyrae stars. In this diagram, we see that both Oosterhoff groups have very similar apocentric distance distributions. We also noticed a sudden decrease at $\approx$\,3\,kpc (log($r_{\text{apo}} \approx$ 0.5\,kpc), which is roughly in agreement with the overlap between bulge, metal-weak thick disk, and inner halo \citep[3.5\,kpc;][]{Ness2013b}. In fact, the majority of the RR~Lyrae stars in our sample stay inside the bulge volume during their full orbits.

Based on the negligible difference in the median value of $z_{\text{max}}$ between the two Oosterhoff groups, we explored whether the Oo\,II variables could follow higher orbits from the Galactic plane and thus possibly come from the halo. Based on the distribution of apocentric distances (Fig.~\ref{fig:ApocHist}), as a nominal boundary for the Galactic bulge we chose $z_{\text{max}} > 3$\,kpc (roughly above $\pm$ 20\,deg at the distance of the Galactic bulge) and calculated the number of stars that fulfill this condition in at least 50\,\% of the iterations of the Monte Carlo simulation. In total, only 34 RR~Lyrae stars fulfill the aforementioned condition. Among these 34 variables, 35\,\% belong to the Oo\,II population. This is a significant increase in comparison with the overall bulge and halo Oosterhoff population where the fraction of Oo\,II pulsators is around 25\,\% \citep{Drake2013,Sesar2013,Prudil2019}. Furthermore, a similar difference prevails even when we apply the less conservative threshold of $z_{\text{max}} > 2$\,kpc resulting in 8\,\% variables that exit the bulge volume on their orbits as members of the Oo\,II component. Therefore, if we consider stars with $z_{\text{max}} > 3.0$\,kpc as halo interlopers passing through the Galactic bulge, their contribution to the Galactic bulge RR~Lyrae population would be $\sim$ 8\,\%, which is roughly in agreement with the 6\,\% estimate by \citet{Kunder2015}. 

\begin{figure} 
\includegraphics[width=\columnwidth]{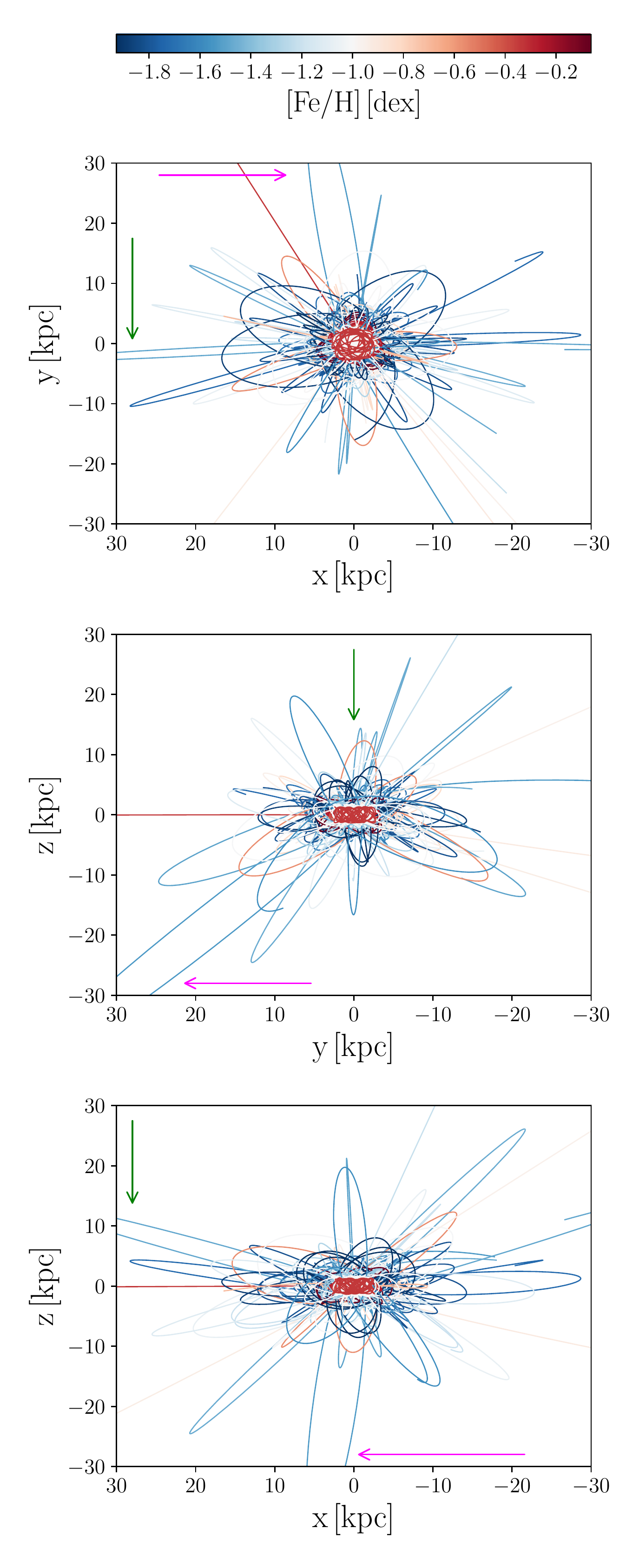}
\caption{The Cartesian spatial distributions with outlined orbital movement color-coded according to the metallicities of individual RR~Lyrae stars, integrated over 1\,Gyr. The arrows mark the high-velocity stars \textit{07001} (magenta arrow) and \textit{09983} (green arrow), disscussed in Sec.~\ref{subsec:Pecul}.}
\label{fig:SpatFeH}
\end{figure}

To further analyze the stars with $z_{\text{max}} > 3.0$\,kpc, we used metallicity estimates for our sample of stars. The spatial distribution of these stars in Galactocentric coordinates with color-coding according to metallicity is shown in Fig.~\ref{fig:SpatFeH}. Stars with high metallicity (blue lines) are more concentrated around low $z_{\text{max}}$ values, while metal-poor variables (red lines) reach higher $z_{\text{max}}$. 

\begin{figure}
\includegraphics[width=\columnwidth]{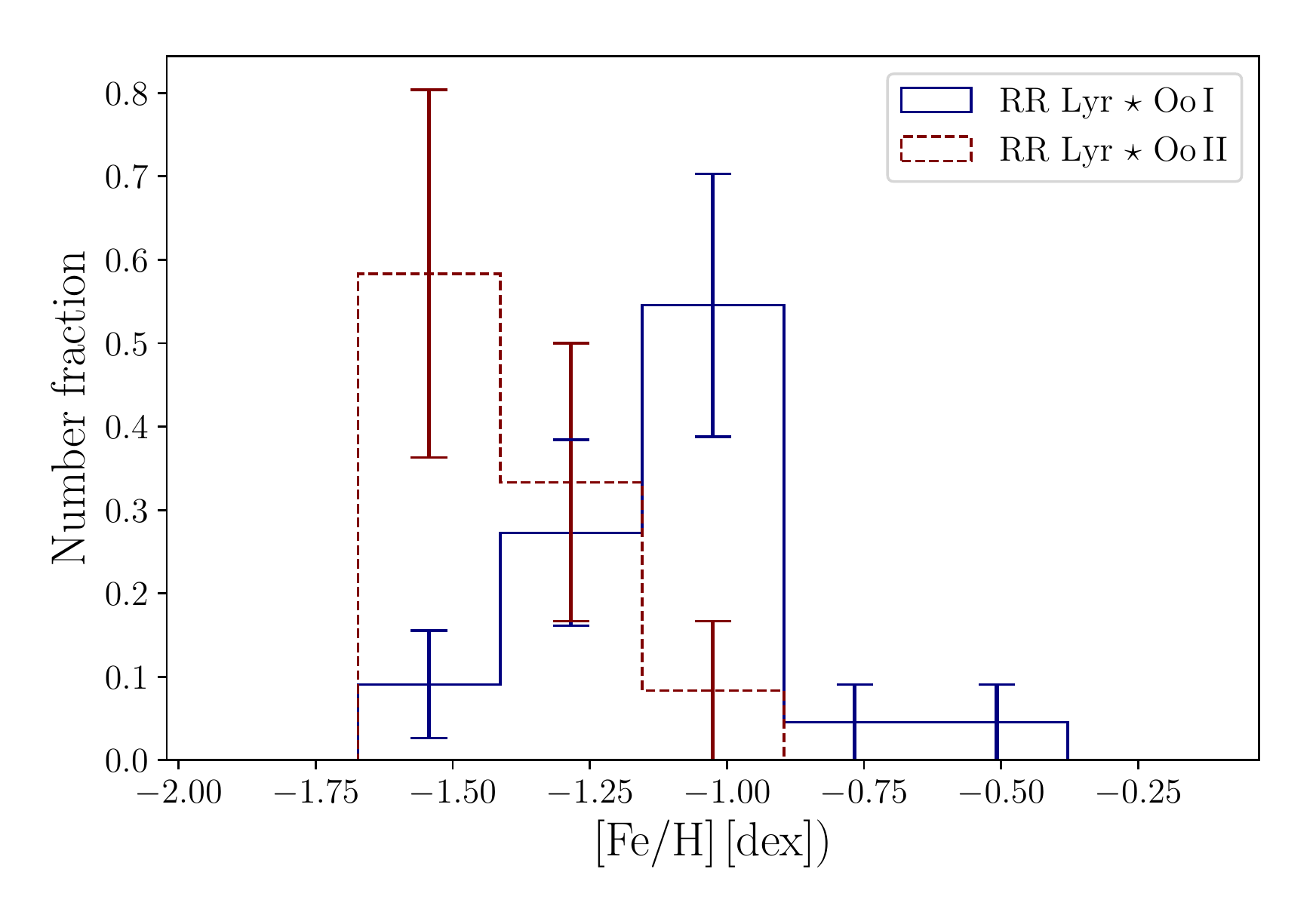} \\
\caption{The metallicity distribution for variables with $z_{\text{max}} > 3.0$\,kpc. The blue and red histograms represent the Oo\,I and Oo\,II groups, respectively.}
\label{fig:SpatFeHzadek}
\end{figure}

In Fig.~\ref{fig:SpatFeHzadek} we show the metallicity distribution of probable halo interlopers. We see a strong peak at $-1.5$\,dex for the Oo\,II group suggesting that these stars probably do not pertain to the Galactic bulge but are currently looping through it, thus contributing to the Oo\,II population in the bulge. The median metallicity of Oo\,I stars in our sample exceeding the $z_{\text{max}} \geq 3.0$\,kpc is $-1.09$\,dex (in comparison, the median metallicity of the entire Oo\,I sample is $-1.04$\,dex), while the median metallicity of Oo\,II is $-1.47$\,dex, which exceeds the $z_{\text{max}} \geq 3.0$\,kpc (in contrast, the median metallicity of the entire Oo\,II population in our sample is $-1.11$\,dex). Here we see a discrepancy between the metallicities of the two Oosterhoff groups in our sample, and between stars whose orbits exceed 3.0\,kpc from the Galactic plane. The metallicity of the Oo\,I group seems to be nearly the same in both cases, and is consistent with bulge values, but the Oo\,II population shows a lower median metallicity that is comparable to that seen in halo stars \citep[$\sim-1.5$\,dex,][]{Ivezic2008,Sesar2011}. 

\subsection{Orbital periods}
We estimated the orbital periods of our studied stars in the following way: First, we calculated 200 times the orbital solution for each star forward in time up to 3\,Gyr, while varying the input parameters with respect to their uncertainties (assuming they have a normal distribution). We recorded the changes in radial distance $R$ over this time span and used the Lomb-Scargle periodogram \citep{Lomb1976,Scargle1982} to calculate the orbital period. The Lomb-Scargle periodograms are used for the analysis of time-series data to search for a potential periodic signal. In the frequency spectrum generated for each orbital solution, we looked for the highest peak and adopted its inverse value as the orbital period $P_{\rm{orbit}}$. 

\begin{figure}
\includegraphics[width=\columnwidth]{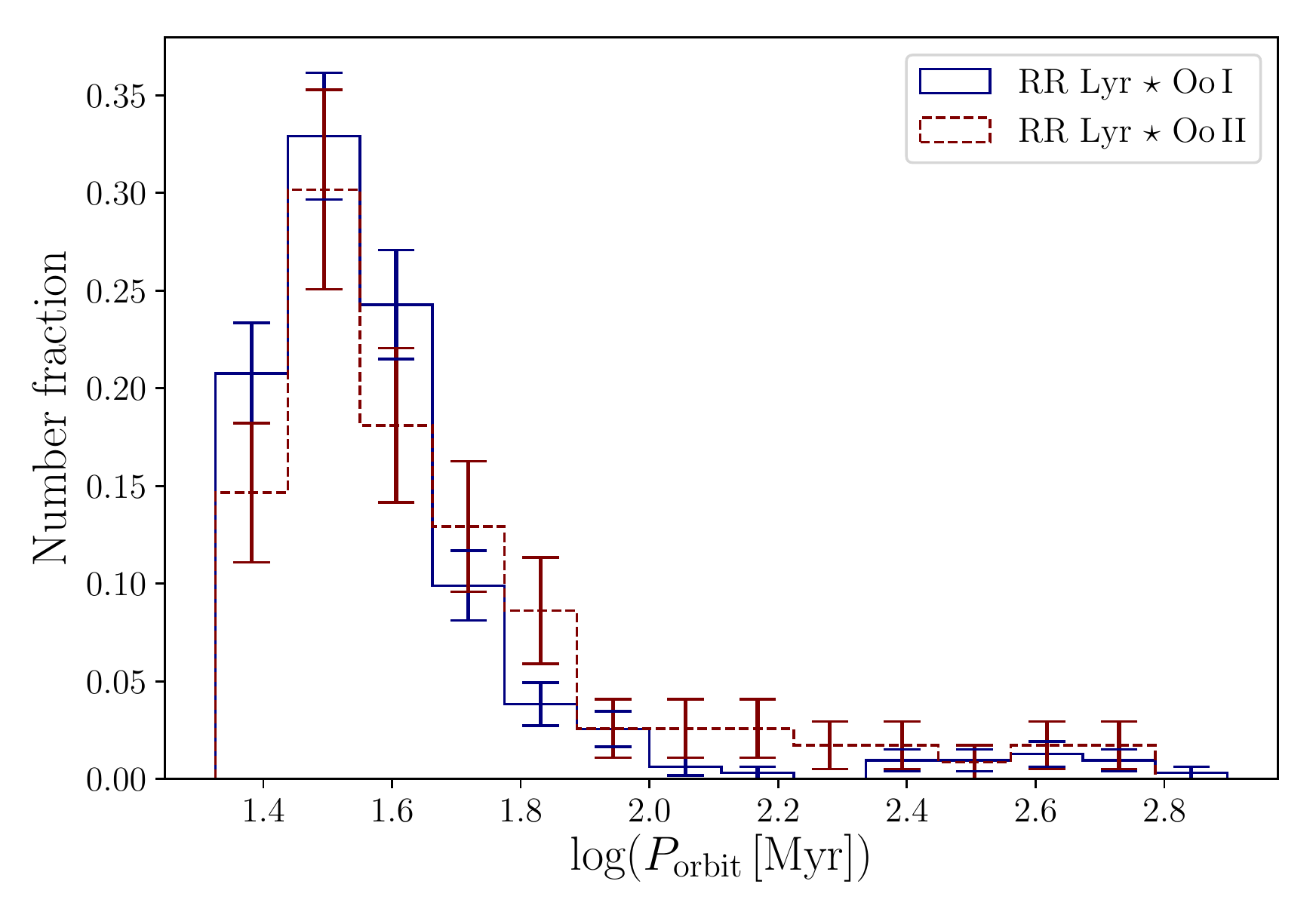} \\
\caption{The orbital period distribution for the studied variables. The blue and red histograms represent the Oo\,I and Oo\,II groups, respectively.}
\label{fig:PeriodyHist}
\end{figure}

Fig.~\ref{fig:PeriodyHist} shows the distribution of the resulting orbital periods for both Oosterhoff populations. The two Oosterhoff groups share a similar median orbital period ($\sim$ 36\,Myr). The shortest orbital periods are only around 21\,Myr. Both distributions appear to be fairly similar on the short orbital period end. We do notice that the fraction of stars in the Oo\,II group increases toward longer orbital periods. At the long end of the orbital periods (over 100\,Myr), 46\,\% of the RR~Lyrae stars belong to the Oo\,II population, which is a significant increase with respect to their fraction in the entire bulge population (25\,\%). This is in agreement with the findings in Figs.~\ref{fig:SpatFeH} and \ref{fig:SpatFeHzadek} where the percentage of Oo\,II stars and fraction of metal-poor variables in general increase with distance from the Galactic center. 

\subsection{Notes on peculiar objects} \label{subsec:Pecul}

Several stars from our sample exhibit peculiar orbits on which they pass through the bulge with high space velocities and have a long orbital period. In this subsection, we will look closely at two such examples.

Almost all of the studied stars in the two top panels of Fig.~\ref{fig:SpatFeH} reach their apocenters at 30\,kpc. One exception is OGLE-BLG-RRLYR-07001 (from hereon \textit{07001}), with $z_{\text{max}}$ above 46\,kpc which is two times larger than the second-highest distance from the Galactic plane in our sample. This star has a typical value of distance (located in the Galactic bulge at 9.5\,kpc), radial velocity, proper motion component $\mu_{\delta}$, and belongs to the Oo\,I group. On the other hand, the $\alpha$ component of its proper motion, $\mu_{\alpha\ast}=9.35\pm0.38$\,mas\,yr$^{-1}$, stands out. Furthermore, its metallicity of [Fe/H]$=-1.37\pm0.19$\,dex is rather peculiar in comparison with the general metallicity of Oo\,I stars in the Galactic bulge. The trajectory in the Galactocentric coordinates of this star can be seen in the top and bottom panels of Fig.~\ref{fig:SpatFeH} as an almost straight line going through $x=0$ on the horizontal axis (see the black arrows in Fig.~\ref{fig:SpatFeH}). The orbital period of \textit{07001} is above 700\,Myr, which puts it in the top 5 highest orbital periods in our sample. This RR~Lyrae star resides in the Galactic bulge for only a few million years. A closer look at the components of the space velocity shows an ordinary velocity component $U=-46\pm10$\,km\,s$^{-1}$, but the $V$ and $W$ components stand out and amount to $278\pm15$\,km\,s$^{-1}$ and $-458\pm18$\,km\,s$^{-1}$, respectively. The true space velocity of \textit{07001} is the second highest among the studied stars, with $sv=538\pm18$\,km\,s$^{-1}$. 

Another interesting object is OGLE-BLG-RRLYR-09983 (from hereon \textit{09983}). With $sv = 542\pm10$\,km\,s$^{-1}$ it has the highest space velocity, and one of the longest orbital periods, $P_{\rm{orbit}} \approx 0.6$\,Gyr. The astrometric properties of this star do not stand out as clearly when compared with the whole sample, but its radial velocity ($-508$\,km\,s$^{-1}$) is the highest within our sample. Such a high radial velocity transforms into the space velocity component $U$ as $486\pm10$\,km\,s$^{-1}$, which makes it the fastest in this component. Unlike \textit{07001}, \textit{09983} belongs to the Oo\,II group, having one of the longest pulsation periods ($>$ 0.81\,day), and a low metallicity, [Fe/H]$=-1.34\pm0.27$\,dex. In addition, its orbit is different from that of \textit{07001} with respect to the Galactic plane. Star \textit{09983} does not deviate from the plane by more than 15\,kpc, but its apocenter lies more than 40\,kpc from the Galactic center. Therefore, it crosses the inner bulge and then for most of its orbit stays fairly close to the Galactic disk in comparison with \textit{07001} with a small excursion in the Galactic inner halo.

For a comparison, the high-velocity RR~Lyrae variable, OGLE-BLG-RRLYR-10353, studied by \citet{Kunder2015} and \citet{Hansen2016}, has $sv=-482\pm22$\,km\,s$^{-1}$, with a somewhat low metallicity \cite[-1.32$\pm$0.19\,dex from its $I$-band light curve\footnote{Using the same procedure as described in \cite{Prudil2019}.} and -0.9$\pm$0.2\,dex from spectroscopy;][]{Hansen2016}. It is necessary to emphasize that this star has been marked as modulated \citep{Prudil2017} and thus might not be suitable for metallicity determination from the photometric data. We note that the escape velocity at the radius of the Galactic bulge is over 600\,km\,s$^{-1}$, thus \textit{07001} and \textit{09983} are still be confined to the MW, but based on the orbital integration they spend the majority of their orbit in halo/disk.

\section{Summary and Conclusions} \label{sec:Conclus}

In this paper, we used RR~Lyrae stars with distances from \citet{Prudil2019}, together with kinematical information from the {\it Gaia} DR2 \citep{Gaia2018} and radial velocities from \citet{Kunder2016} to study possible differences in their kinematical properties per Oosterhoff group in the Galactic bulge. For the 429 variables in our sample, we calculated 6D kinematical solution and predicted their Galactic orbits \citep{galpy2015}.

The RR~Lyrae stars appear to be part of the kinematically hot component of the Galactic bulge based on their tangential velocities, similar to the Galactic bulge Type II Cepheids \citep{Braga2018}. The orbits of the majority of the studied RR~Lyrae stars are fully confined to the bulge volume, with only 8\,\% of the sample (34 stars) have their apocenters more than 3\,kpc from the Galactic Center. This is in agreement with the 6\,\% estimate by \citet{Kunder2015}. Based on the orbital parameters both Oosterhoff groups seem fairly similar, with the exception of $z_{\text{max}}$, where we see an increase in the Oo\,II population as we move away from the Galactic plane. There more than one-third belong the Oo\,II population, which is in contrast with 25\,\% ratio found by \citet{Prudil2019} for the Galactic bulge.

We identified six high-velocity RR~Lyrae stars that have higher space velocity than RR~Lyrae variable studied by \citet{Kunder2015} and \citet{Hansen2016}. Neither of their space velocities exceeds the escape velocity at the radius of the Galactic bulge. Nevertheless, they can serve as targers for future spectroscopic studies as probes of the Galactic bulge and halo.

In summary, the kinematics of bulge RR~Lyrae stars belonging to different Oosterhoff groups is fairly similar. A small part of our sample probably consists of halo interlopers, also from a chemical perspective. Based on the kinematics and orbital solutions for our stars, the vast majority do not seem to contribute to the boxy/peanut component consisting of the bulge's intermediate-age population \citep{McWilliam2010,Wegg2013}; but rather they seem to be part of a primordial component of the Galactic bulge, confirming the previous findings by \citet{Dekany2013,Kunder2016}.

\section*{Acknowledgements}

This work has made use of data from the European Space Agency (ESA) mission {\it Gaia} (\url{https://www.cosmos.esa.int/gaia}), processed by the {\it Gaia} Data Processing and Analysis Consortium (DPAC, \url{https://www.cosmos.esa.int/web/gaia/dpac/consortium}). Funding for the DPAC has been provided by national institutions, in particular the institutions participating in the {\it Gaia} Multilateral Agreement. Z.P. is grateful to A. Kunder for useful comments on the manuscript. Z.P. acknowledges the support of the Hector Fellow Academy. E.K.G and I.D. were supported by Sonderforschungsbereich SFB 881 ``The Milky Way System'' (subprojects A02, A03, A11) of the German Research Foundation (DFG). Support for M.C. is provided by Fondecyt through grant \#1171273; the Ministry for the Economy, Development, and Tourism's Millennium Science Initiative through grant IC\,120009, awarded to the Millennium Institute of Astrophysics (MAS); by Proyecto Basal AFB-170002; and by CONICYT's PCI program through grant DPI20140066. M.S. acknowledges support from Postdoc$@$MUNI project CZ.02.2.69/0.0/0.0/16$\_$027/0008360 and GACR international grant 17-01752J. R.S. was supported by the National Science Center, Poland, grant agreement DEC-2015/17/B/ST9/03421.



\newpage 


\bsp	
\label{lastpage}

\begin{thebibliography}{99} 
\bibitem[\protect\citeauthoryear{Alcock et al.}{1997}]{Alcock1997} Alcock C., et al., 1997, ApJ, 474, 217 
\bibitem[\protect\citeauthoryear{Bland-Hawthorn \& Gerhard}{2016}]{BHG2016} Bland-Hawthorn J., Gerhard O., 2016, ARA\&A, 54, 529 
\bibitem[\protect\citeauthoryear{Barbuy, Chiappini, \& Gerhard}{2018}]{Barbuy2018} Barbuy B., Chiappini C., Gerhard O., 2018, ARA\&A, 56, 223 
\bibitem[\protect\citeauthoryear{Bovy et al.}{2012}]{Bovy2012} Bovy J., et al., 2012, ApJ, 759, 131 
\bibitem[\protect\citeauthoryear{Bovy}{2015}]{galpy2015} Bovy J., 2015, ApJS, 216, 29 
\bibitem[\protect\citeauthoryear{Braga et al.}{2018}]{Braga2018} Braga V.~F., Bhardwaj A., Contreras Ramos R., Minniti D., Bono G., de Grijs R., Minniti J.~H., Rejkuba M., 2018, A\&A, 619, A51
\bibitem[\protect\citeauthoryear{Catelan}{2009}]{Catelan2009} Catelan M., 2009, Ap\&SS, 320, 261 
\bibitem[\protect\citeauthoryear{Catelan \& Smith}{2015}]{Catelan2015Book} Catelan M., Smith H.~A., 2015, Pulsating Stars, Wiley, New York 
\bibitem[\protect\citeauthoryear{Clementini et al.}{2005}]{Clementini2005} Clementini G., Gratton R.~G., Bragaglia A., Ripepi V., Martinez Fiorenzano A.~F., Held E.~V., Carretta E., 2005, ApJ, 630, L145 
\bibitem[\protect\citeauthoryear{Dehnen}{2000}]{Dehnen2000} Dehnen W., 2000, AJ, 119, 800 
\bibitem[\protect\citeauthoryear{D{\'e}k{\'a}ny et al.}{2013}]{Dekany2013} D{\'e}k{\'a}ny I., Minniti D., Catelan M., Zoccali M., Saito R.~K., Hempel M., Gonzalez O.~A., 2013, ApJ, 776, L19 
\bibitem[\protect\citeauthoryear{D{\'e}k{\'a}ny et al.}{2018}]{Dekany2018} D{\'e}k{\'a}ny I., Hajdu G., Grebel E.~K., Catelan M., Elorrieta F., Eyheramendy S., Majaess D., Jord{\'a}n A., 2018, ApJ, 857, 54 
\bibitem[\protect\citeauthoryear{Drake et al.}{2013}]{Drake2013} Drake A.~J., et al., 2013, ApJ, 763, 32 
\bibitem[\protect\citeauthoryear{Dwek et al.}{1995}]{Dwek1995} Dwek E., et al., 1995, ApJ, 445, 716 
\bibitem[\protect\citeauthoryear{ESA}{1997}]{ESA1997} ESA ed. 1997, The HIPPARCOS and TYCHO catalogues. Astrometric and photometric star catalogues derived from the ESA HIPPARCOS Space Astrometry Mission, 1997, ESA Special Publication, Vol. 1200  
\bibitem[\protect\citeauthoryear{Freeman et al.}{2013}]{Freeman2013} Freeman K., et al., 2013, MNRAS, 428, 3660 
\bibitem[\protect\citeauthoryear{Gaia Collaboration et al.}{2016}]{Gaia2016} Gaia Collaboration, et al., 2016, A\&A, 595, A1 
\bibitem[\protect\citeauthoryear{Gaia Collaboration et al.}{2018}]{Gaia2018} Gaia Collaboration, et al., 2018, A\&A, 616, A1 
\bibitem[\protect\citeauthoryear{Gillessen et al.}{2009}]{Gillessen2009} Gillessen S., Eisenhauer F., Trippe S., Alexander T., Genzel R., Martins F., Ott T., 2009, ApJ, 692, 1075 
\bibitem[\protect\citeauthoryear{Guggenberger et al.}{2012}]{Guggenberger2012} Guggenberger E., et al., 2012, MNRAS, 424, 649 
\bibitem[\protect\citeauthoryear{Hansen et al.}{2016}]{Hansen2016} Hansen C.~J., Rich R.~M., Koch A., Xu S., Kunder A., Ludwig H.-G., 2016, A\&A, 590, A39 
\bibitem[\protect\citeauthoryear{Hartigan \& Hartigan}{1985}]{Hartigan1985} Hartigan, J. A., and P. M. Hartigan. "The Dip Test of Unimodality." The Annals of Statistics, vol. 13, no. 1, 1985, JSTOR, pp. 70.  
\bibitem[\protect\citeauthoryear{Ivezi{\'c} et al.}{2008}]{Ivezic2008} Ivezi{\'c} {\v Z}., et al., 2008, ApJ, 684, 287 
\bibitem[\protect\citeauthoryear{Johnson \& Soderblom}{1987}]{Johnson1987} Johnson D.~R.~H., Soderblom D.~R., 1987, AJ, 93, 864 
\bibitem[\protect\citeauthoryear{Kerr \& Lynden-Bell}{1986}]{Kerr1986} Kerr F.~J., Lynden-Bell D., 1986, MNRAS, 221, 1023 
\bibitem[\protect\citeauthoryear{Layden et al.}{1996}]{Layden1996} Layden A.~C., Hanson R.~B., Hawley S.~L., Klemola A.~R., Hanley C.~J., 1996, AJ, 112, 2110 
\bibitem[\protect\citeauthoryear{Lomb}{1976}]{Lomb1976} Lomb N.~R., 1976, Ap\&SS, 39, 447
\bibitem[\protect\citeauthoryear{Kormendy \& Kennicutt}{2004}]{Kormendy2004} Kormendy J., Kennicutt R.~C., Jr., 2004, ARA\&A, 42, 603 
\bibitem[\protect\citeauthoryear{Kunder et al.}{2012}]{Kunder2012} Kunder A., et al., 2012, AJ, 143, 57 
\bibitem[\protect\citeauthoryear{Kunder et al.}{2015}]{Kunder2015} Kunder A., et al., 2015, ApJ, 808, L12 
\bibitem[\protect\citeauthoryear{Kunder et al.}{2016}]{Kunder2016} Kunder A., et al., 2016, ApJ, 821, L25
\bibitem[\protect\citeauthoryear{Kunder et al.}{2018}]{Kunder2018} Kunder A., et al., 2018, AJ, 155, 171
\bibitem[\protect\citeauthoryear{McWilliam \& Zoccali}{2010}]{McWilliam2010} McWilliam A., Zoccali M., 2010, ApJ, 724, 1491 
\bibitem[\protect\citeauthoryear{Minniti et al.}{2010}]{Minniti2010} Minniti D., et al., 2010, NewA, 15, 433 
\bibitem[\protect\citeauthoryear{Miyamoto \& Nagai}{1975}]{MiyamotoNagai1975} Miyamoto M., Nagai R., 1975, PASJ, 27, 533 
\bibitem[\protect\citeauthoryear{Navarro, Frenk, \& White}{1997}]{Navarro1997} Navarro J.~F., Frenk C.~S., White S.~D.~M., 1997, ApJ, 490, 493 
\bibitem[\protect\citeauthoryear{Ness et al.}{2013a}]{Ness2013a} Ness M., et al., 2013a, MNRAS, 432, 2092 
\bibitem[\protect\citeauthoryear{Ness et al.}{2013b}]{Ness2013b} Ness M., et al., 2013b, MNRAS, 430, 836 
\bibitem[\protect\citeauthoryear{Obreja et al.}{2013}]{Obreja2013} Obreja A., Dom{\'{\i}}nguez-Tenreiro R., Brook C., Mart{\'{\i}}nez-Serrano F.~J., Dom{\'e}nech-Moral M., Serna A., Moll{\'a} M., Stinson G., 2013, ApJ, 763, 26 
\bibitem[\protect\citeauthoryear{Oosterhoff}{1939}]{Oosterhoff1939} Oosterhoff P.~Th., 1939, Obs, 62, 104
\bibitem[\protect\citeauthoryear{Pietrukowicz et al.}{2015}]{Pietrukowicz2015} Pietrukowicz P., et al., 2015, ApJ, 811, 113 
\bibitem[\protect\citeauthoryear{Pritzl et al.}{2000}]{Pritzl2000} Pritzl B., Smith H.~A., Catelan M., Sweigart A.~V., 2000, ApJ, 530, L41 
\bibitem[\protect\citeauthoryear{Prudil \& Skarka}{2017}]{Prudil2017} Prudil Z., Skarka M., 2017, MNRAS, 466, 2602 
\bibitem[\protect\citeauthoryear{Prudil et al.}{2019}]{Prudil2019} Prudil Z., D{\'e}k{\'a}ny I., Catelan M., Smolec R., Grebel E.~K., Skarka M., 2019, MNRAS, 484, 4833
\bibitem[\protect\citeauthoryear{Rosenbaum}{2005}]{Rosenbaum2005} Rosenbaum, P., 2005, JRSS B, 67, 515--530
\bibitem[\protect\citeauthoryear{Scargle}{1982}]{Scargle1982} Scargle J.~D., 1982, ApJ, 263, 835 
\bibitem[\protect\citeauthoryear{Sesar, Juri{\'c}, \& Ivezi{\'c}}{2011}]{Sesar2011} Sesar B., Juri{\'c} M., Ivezi{\'c} {\v Z}., 2011, ApJ, 731, 4 
\bibitem[\protect\citeauthoryear{Sesar et al.}{2013}]{Sesar2013} Sesar B., et al., 2013, AJ, 146, 21 
\bibitem[\protect\citeauthoryear{Sch{\"o}nrich, Binney, \& Dehnen}{2010}]{SCHONRICH2010} Sch{\"o}nrich R., Binney J., Dehnen W., 2010, MNRAS, 403, 1829 
\bibitem[\protect\citeauthoryear{Sch{\"o}nrich}{2012}]{SCHONRICH2012} Sch{\"o}nrich R., 2012, MNRAS, 427, 274 
\bibitem[\protect\citeauthoryear{Smolec et al.}{2016}]{Smolec2016} Smolec R., Prudil Z., Skarka M., Bakowska K., 2016, MNRAS, 461, 2934 
\bibitem[\protect\citeauthoryear{Soszy{\'n}ski et al.}{2017}]{Soszynski2017} Soszy{\'n}ski I., et al., 2017, AcA, 67, 297 
\bibitem[\protect\citeauthoryear{Udalski, Szyma{\'n}ski, \& Szyma{\'n}ski}{2015}]{Udalski2015} Udalski A., Szyma{\'n}ski M.~K., Szyma{\'n}ski G., 2015, AcA, 65, 1
\bibitem[\protect\citeauthoryear{Vasiliev}{2019}]{Vasiliev2019} Vasiliev E., 2019, MNRAS, 484, 2832 
\bibitem[\protect\citeauthoryear{Wegg \& Gerhard}{2013}]{Wegg2013} Wegg C., Gerhard O., 2013, MNRAS, 435, 1874 
\bibitem[\protect\citeauthoryear{Weiland et al.}{1994}]{Weiland1994} Weiland J.~L., et al., 1994, ApJ, 425, L81 
\bibitem[\protect\citeauthoryear{Wyse, Gilmore, \& Franx}{1997}]{Wyse1997} Wyse R.~F.~G., Gilmore G., Franx M., 1997, ARA\&A, 35, 637 
\bibitem[\protect\citeauthoryear{Zoccali et al.}{2008}]{Zoccali2008} Zoccali M., Hill V., Lecureur A., Barbuy B., Renzini A., Minniti D., G{\'o}mez A., Ortolani S., 2008, A\&A, 486, 177 
\end{thebibliography}
\end{document}